\documentclass[aps,prl,twocolumn,a4paper,floatfix,showpacs]{revtex4-1} 
\usepackage{amsmath}
\usepackage{amsfonts}
\usepackage{amssymb}
\usepackage{graphicx}
\graphicspath{{./}}
\usepackage{epstopdf}
\begin{document}
\title{Anderson localization in an interacting fermionic system}

\author{Francesco Massel}

\affiliation{University of Helsinki, Department of Mathematics and
  Statistics, P.O. Box 68, FIN-00014, Finland}

\begin{abstract}
  In the present article, we discuss the role played by the
  interaction in the Anderson localization problem, for a system of
  interacting fermions in a one-dimensional disordered lattice,
  described by the Fermi Hubbard Hamiltonian, in presence of an
  on-site random potential. We show that, given the proper
  identification of the elementary excitations of the system described
  in terms of doublons and unpaired particles, the Anderson
  localization picture survives.  Ensuing a ``global quench'', we show
  that the system exhibits a rich localization scenario, which can be
  ascribed to the nearly-free dynamics of the elementary excitations
  of the Hubbard Hamiltonian. 
\end{abstract}
\maketitle

Anderson localization (AL) represents one of the most prominent
manifestations of the role of disorder in quantum mechanical systems
\cite{ANDERSON:1958uc}. Over the last 5 decades, the concept of AL has
been successfully employed to describe the physics of localization in
the most diverse contexts, ranging from the localization of optical
\cite{Wiersma:1997vo,Schwartz:2007co} and acoustical waves
\cite{Hu:2008jy} to that of matter waves in solid-state systems
\cite{LEE:1985tl} and ultracold gases
\cite{Billy:2008gs,Roati:2008hi,Deissler:2010hl}.

In its original formulation, AL predicts the absence of diffusion and
exponential localization of a quantum particle due to coherent
multiple scatterings with a random potential
\cite{ANDERSON:1958uc}. This picture allows the description of the
role of disorder in a many-body situation when the interactions among
constituents are absent, and thus the physics can be recast in terms
of a one-body problem. Conversely, the description of the physics of
localization in a genuinely many-body situation -- i.e.  when two-body
interactions are present-- has proven to be an outstanding challenge
(see e.g. \cite{SanchezPalencia:2010db}), with different approaches
providing different scenarios for the effect of interactions on
localization
\cite{SanchezPalencia:2007ca,Paul:2007jw,Pikovsky:2008bd,Kopidakis:2008cb,Deissler:2010hl,Lugan:2011kda,Delande:2013gh}.

In the recent past ultracold atomic gases have become the ideal
test-bed for models in condensed-matter physics owing to the wide
tunability of their characteristic parameters, such as the possibility
of controlling the particle-particle interaction and the system
dimensionality. They thus represent optimal candidates for the
investigation of AL, as attested by the experimental observation of AL
in a Bose-Einstein condensate of non-interacting alkali atoms
\cite{Billy:2008gs,Roati:2008hi}. These two experiments have provided
evidence for AL in matter waves while in \cite{Deissler:2010hl}, the
effect of a weak interacting potential on the localization scenario
for a bosonic system was explored.  Moreover, the effect of
interaction has been recently investigated numerically and
theoretically in \cite{Delande:2013gh} , showing the persistence of AL
in a one-dimensional bosonic system in presence of attractive
interaction beyond a mean-field description.

In this article we theoretically and numerically study a ``global
quantum quench'' \cite{Calabrese:2006bg} of an interacting fermionic
cloud of ultracold atoms in a disordered lattice%, pointing out the
%role of two-body interactions in the localization scenario
. Our investigation is focused on the dynamics generated by the
one-dimensional Fermi-Hubbard model in presence of an on-site
disordered potential on a two-component fermionic gas.

The relevance of this model in the context of the study of the
combined effect of interaction and disorder lies in the fact that the
Hubbard model constitutes the prototypical model for the study of
interactions in a solid-state framework. Furthermore, our analysis is
closely related to the current experimental trends in the field,
represented by the experiments aimed at establishing the AL picture
for non-interacting bosonic systems \cite{Billy:2008gs,Roati:2008hi},
or by the expansion of a fermionic BI in an otherwise empty lattice in
two spatial dimensions (in absence of disorder)
\cite{Schneider:2012ke}. Since the setup here considered represents a
natural extension to disordered systems of the one considered in
\cite{Kajala:2011ho}, we expect that it can be realized experimentally
in the near future.

On general grounds, beyond any specific reference to a given model, we
here point out how the correct identification of the elementary
excitations within any genuinely many-body system constitutes the
fundamental step in the definition of the framework for the
investigation of the interaction/disorder interplay. In this same
spirit, in \cite{Lugan:2011kda}, the localization properties of
Bogoliubov quasiparticles in an interacting Bose gas was studied. Our
goal is to illustrate here how AL survives for a fermionic system in
presence of two-body interactions ---either attractive or repulsive--
and how the rich localization scenario exhibited by the system is
associated to the collective character of its excitations.

Our numerical simulations are performed using a time-evolving block
decimation algorithm (TEBD), which represents an essentially exact
numerical technique for the simulation of the ground-state and the
dynamical properties of one-dimensional quantum lattice systems
\cite{Vidal:2004jc}. In addition, the existence of the Bethe-ansatz
solution for the 1D Hubbard chain \cite{Essler:2005uw} in absence of
disorder, represents our starting point for the theoretical discussion
about the disorder and interaction beyond mean-field and perturbative
approaches.

\begin{figure}
 \includegraphics[width=0.45\textwidth]{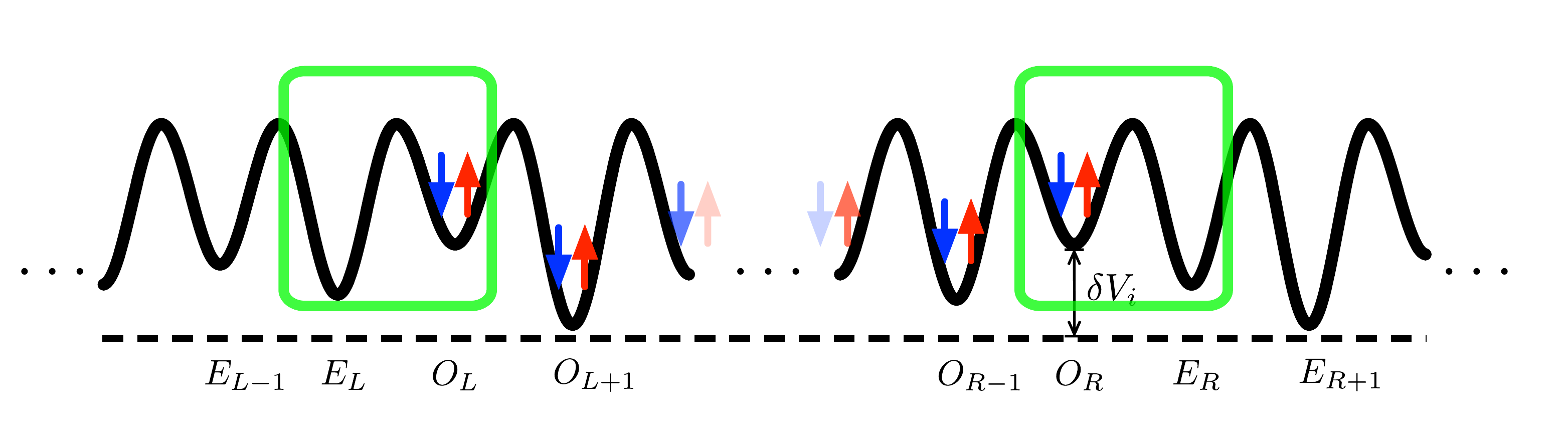}
 \caption{In the initial configuration the central portion of the
   lattice is filled with two atoms per site. At $t=0^+$ the particles
   are allowed to expand in an empty lattice, whose on-site potential
   is a random variable, uniformly distributed between $0$ and
   $V_0$. Sites $E_L$, $O_L$ and $O_R$, $E_R$ represent the left and
   right edges of the cloud respectively.}
  \label{fig:1}
\end{figure}

In our analysis, we consider a system of
$N=N_\uparrow+N_\downarrow=40$ fermionic atoms with equal pseudospin
populations ($N_\uparrow=N_\downarrow=20$) in a one-dimensional
lattice with $L=256$ sites.
% whose central $20$ sites are occupied by
%one atom per pseudospin species, i.e. $\langle
%n_{i\,\uparrow}\rangle=\langle n_{i\,\downarrow}\rangle=1$ if $119
%\leq i \leq 138 $, and $0$ otherwise.
  The system is prepared at
$t=0^-$ by loading a band insulator (BI) state in the central portion
of the disordered lattice (see Fig. \ref{fig:1}).
At $t=0^+$ the BI is allowed to expand in the empty lattice.  The
dynamics is described in terms of the Fermi-Hubbard Hamiltonian with a
disordered on-site potential
\begin{equation}
    \label{eq:hubb}
    H=H_{J}+H_{U}+\sum_{i=1}^{L}V_i \left(n_{i\uparrow}+n_{i\downarrow}\right),
  \end{equation}
  where $H_{J}=-J\sum_{i=1,\sigma=\uparrow\,\downarrow
  }^{L}c_{i\sigma}^{\dagger}c_{i+1\sigma}+h.c.$,
  $H_{U}=U\sum_{i=1}^{L}n_{i\uparrow}n_{i\downarrow}$, $J$ is the
  hopping amplitude, $U$ is the on-site interaction, $V_i$ is the
  random on-site potential with probability ditribution $P(V_i)$. In
  the following, we will focus on the case where $V_i$ is uniformly
  distributed between $0$ and $V_0$. We expect that a different choice
  for the disorder distribution will alter the specific value of the
  localization length --see Eq. \eqref{eq:1}--, but the qualitative
  content of our predictions is expected to be unchanged.
 
\begin{figure}[!ht]
 \includegraphics[angle=-90,width=0.5\textwidth]{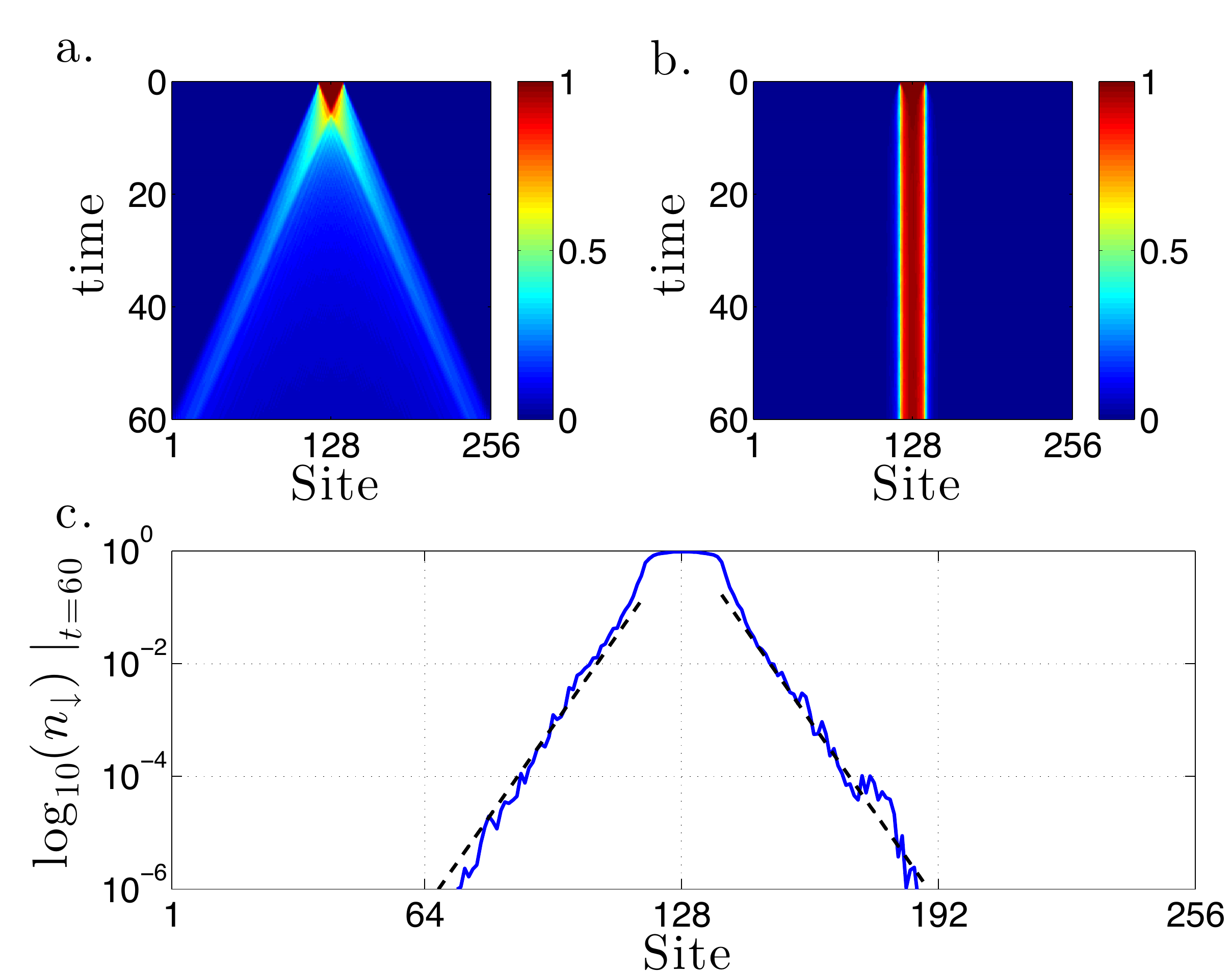}
 \caption{a. Expectation value of the number of particles $\langle
   n_{i\,\sigma}\rangle(t)$ with spin $\sigma=\uparrow\,\downarrow$,
   for $U=0$, and $V_0=0$. It is possible to see how the particle
   wavefront propagates with velocity $\|v\|=2J$. b. $\langle
   n_{i\,\sigma}\rangle(t)$ for $U=0$ and $V_0=5$. Particles are
   exponentially localized. c. Logarithmic plot of $\langle
   n_{i\,\sigma}\rangle(t)\left.\right|_{t=60 J^{-1}}$ (blue curve),
     compared with the theoretical value the exponential decay over
     the localization length $l_{loc}$ given by Eq. \eqref{eq:1}
     (black dotted curve).}
\label{fig:2}
\end{figure}

\begin{figure}
 \includegraphics[width=0.5\textwidth]{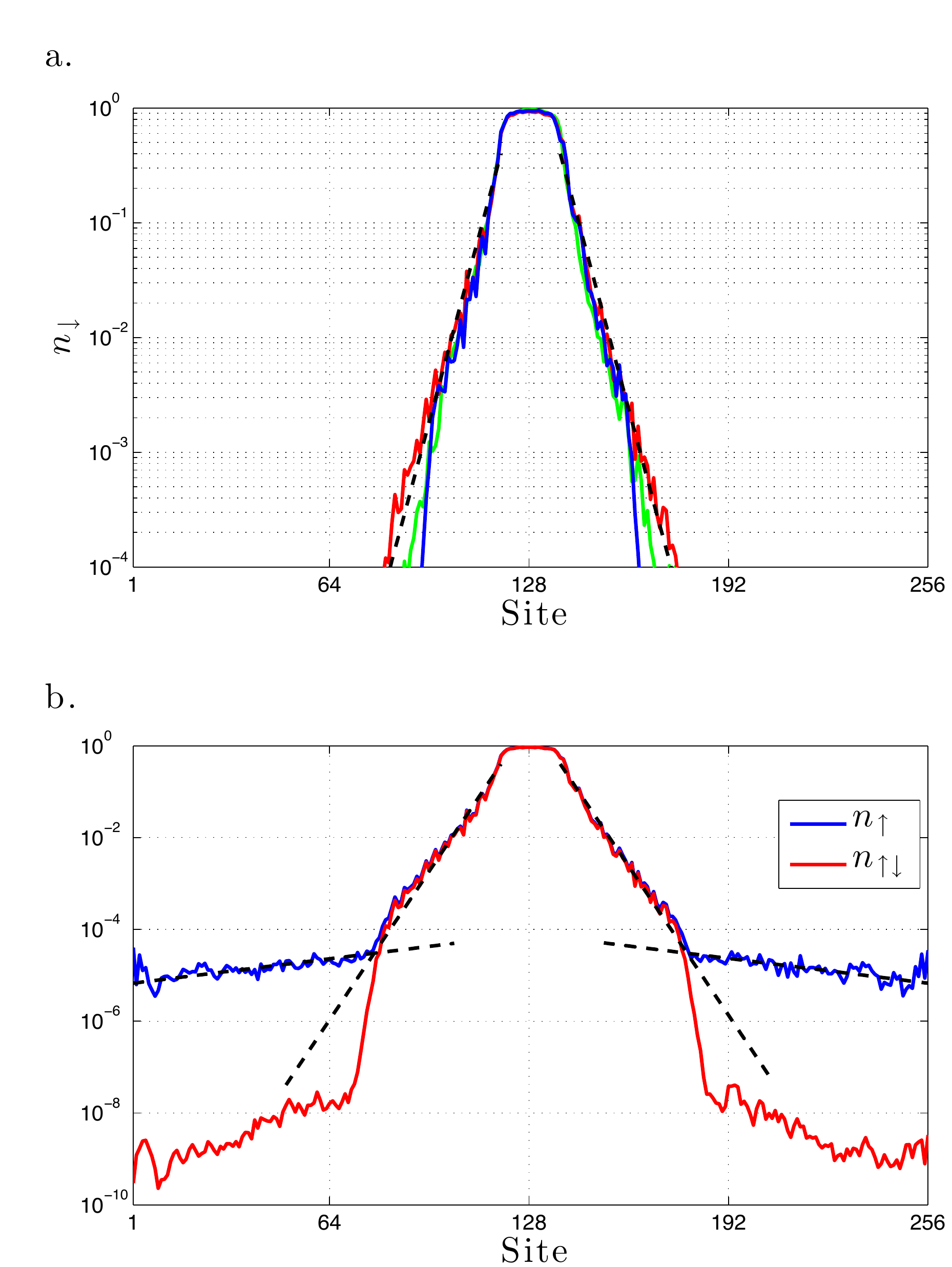}
 \caption{a. Numerical value of $\langle n_{i\,\sigma}\rangle(t)$
   ($U=0$, $V_0/J=5$, green curve; $U/J=-5$, $V_0/J=1$, red curve;
   $U/J=-10$,$V_0/J=0.5$, blue curve), compared with the theoretical
   result (black curve). In agreement with the two-fluid description,
   the three set of parameters exhibit the same localization
   length. b. Numerical value of $\langle n_{i\,\sigma}\rangle(t^*)$
   (blue curve) and $\langle n_{i\,\sigma}\rangle(t^*)$ ($U/J=-5$,
   $V_0/J=1$, $t^*=60 J^{-1}$) compared with the theoretical value of
   the localization length for doublons and unpaired particles. The
   central core exhibits a localization length characteristic of
   doublons localization, while the outer wings exhibit a localization
   length corresponding to the localization of unpaired particles.}
  \label{fig:3}
\end{figure}

\begin{figure}
\includegraphics[width=0.45\textwidth]{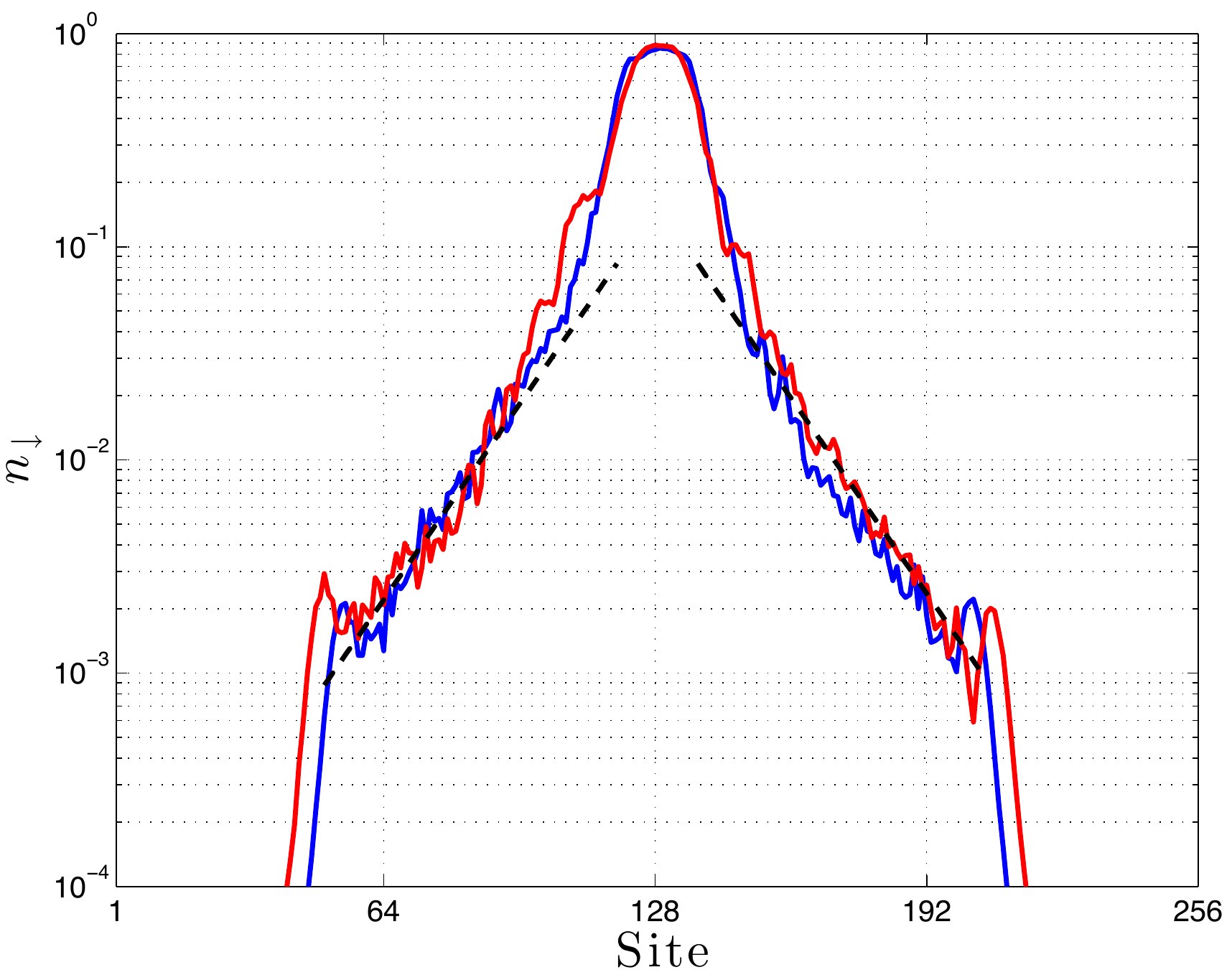}
 \caption{BI expansion for $U=-5$, $V_0=0.4$ (blue curve), compared
   with the expansion for $U=5$, $V_0=0.4$ (red curve). Both curves
   exhibit the same exponential decay in agreement with the theory
   (black dashed line).}
  \label{fig:4}
\end{figure}

  In all our simulations we have considered a Schmidt number
  $\chi=160$, corresponding, in the worst-case scenario ($U/J=-5 $,
  $V_0/J= 1$), to a discarded weight of $10^{-5}$ \cite{Vidal:2004jc}.
  The results depicted in Figs. \ref{fig:2}-\ref{fig:4} are obtained
  averaging over 100 disorder realizations for the $U=0$ case and 10
  disorder realizations for the interacting cases.  The crucial aspect
  allowing reliable simulations with the TEBD algorithm in our system,
  is that the entropy of entanglement, describing the degree of
  entanglement between left-right partitions of the chain does not
  grow when Anderson localisation sets in \cite{Delande:2013gh}. The
  explanation of this phenomenon in our setup cannot be ascribed to
  the low-energy character of the many-body state as in
  \cite{Delande:2013gh}, since the initial BI state represents a
  highly excited state for the system. This saturation is in
  disagreement with the results obtained in \cite{Bardarson:2012gc}
  for the XXZ model, where the nearest-neighbour character of the
  interaction -- as opposed to the on-site nature of the interaction
  in the Hubbard model-- might lead to a different localization
  scenario, and to a different behaviour of the entropy of
  entanglement.

  Firstly, we have compared, for a non-interacting system ($U=0$), the
  expansion in a lattice where disorder is absent (Fig. \ref{fig:2}a.)
  with the expansion in presence of uniform disorder ($V_0=5 J$,
  Fig. \ref{fig:2}b.). In the case of a non-interacting system, the
  presence of an on-site disordered potential prevents the fermionic
  cloud from expanding as it does in the case where disorder is absent,
  confirming the AL picture for non-interacting atoms
  \cite{ANDERSON:1958uc}.  Following \cite{Izrailev:1999uc}, for
  non-interacting atoms in presence of uniformly distributed disorder,
  the localisation length of an expanding cloud, initially prepared in
  a BI state is given by
  \begin{equation}
    \label{eq:1}
    l_{loc}^{-1}(V_0/J)=\frac{1}{2}\ln\left[1+\frac{V_0^2/J^2}{16}\right]+
                    \frac{4\arctan\left[V_0/4J\right]}{V_0/J}-1,
  \end{equation}
  which represents the localization length of a particle lying at the
  band centre. It is possible to show that the overall localization
  length of the system of non-interacting particles is given by this
  length (see Supplementary Information).

  The logarithmic plot of the particle distribution $\langle
  n_{i\,\sigma} \rangle$ ($\sigma=\uparrow,\,\downarrow$) at time $t=
  60 J^{-1}$, given in Fig. \ref{fig:3}b. , exhibit an exponential
  decay compatible with a localisation length expressed by
  Eq. \eqref{eq:1}, confirming the expected effect of an on-site
  disordered potential.
 
In order to investigate the effect of disorder in the expansion for
the interacting case, we first observe that, in absence of
interaction, the dynamics of the expansion can be described in terms
of a two-fluid model \cite{Kajala:2011ho}. This description is
obtained from the Bethe-ansatz solution of the Fermi-Hubbard
Hamiltonian \cite{Massel:2013ge,Kajala:2011ho,Essler:2005uw}, which
allows a strong-coupling limit description of the collective
excitations of the system in terms of unpaired particles and local
pairs (\textit{doublons}), whose local density is given by
$n^{d}_i=\langle \hat{n}_{i\,\uparrow}\hat{n}_{i\,\downarrow}
\rangle$, and $n^u_{i,\sigma}=\langle \hat{n}_{i\,\sigma}
\rangle-\langle \hat{n}_{i\,\uparrow}\hat{n}_{i\,\downarrow} \rangle$
respectively.

The description of the elementary excitations in the one-dimensional
Hubbard model in terms of doublons and unpaired particles can be
obtained considering the string hypothesis for the solution of the
Lieb-Wu equations: in this framework, the doublons correspond to
$k-\Lambda$ strings, while the single $k_j$s correspond to unpaired
particles (see Supplementary Information)
\cite{Massel:2013ge,Essler:2005uw}. Following the string hypothesis,
in the strong-coupling limit, the dispersion relation for unpaired
particles and doublons read respectively
\begin{align}
  \label{eq:4}
  \epsilon_{u}&=-2J \cos(k_u) \\
\label{eq:5}
  \epsilon_{d}&=-4J^2/\left| U \right| \cos(k_d).
\end{align}
Eqs. (\ref{eq:4},~\ref{eq:5}) correspond to the dispersion relation
for free particles with hopping parameter given respectively by $J^u=J$
and $J^d=2J^2/\left| U \right|$, thus describing the free evolution of a
(non-interacting) gas of unparied particles and doublons.
The dynamics at $t=0^+$ can be intuitively described in terms of a
local two-sites dynamics taking place at the edge of the BI cloud (see
Fig. \ref{fig:1}), this description
is possible owing to the specific form of the initial state: sites
away from the edge do not contribute to the initial dynamics, because
they are either empty or completely Pauli blocked
\cite{Kajala:2011ho}. In this picture the initial state where a
doublon is present on the edge of the cloud, will evolve into a
superposition of a doublon (on either side of the edge of the BI
cloud) and two unpaired particles
\begin{align}
  \nonumber
 U(t) \left|
  \emptyset,\uparrow\,\downarrow \right\rangle= 
    \alpha(t) \left| \emptyset,\uparrow\,\downarrow \right\rangle 
+ \beta(t)   \left| \uparrow\,\downarrow, \emptyset\right\rangle \\
+ \gamma_1(t) \left| \uparrow ,\downarrow\right\rangle 
+ \gamma_2(t) \left| \downarrow, \uparrow \right\rangle 
\label{eq:2}
\end{align}

The ensuing dynamics of the many body-system can be approximately
described as the independent dynamics of these two components,
following the dispersion relation given by Eqs. \eqref{eq:4} and
\eqref{eq:5} for unpaired particles and doublons respectively: and
therefore describing the expansion of two fluids characterized by
hopping parameters $J^u$ and $J^d$.

We are now in the position to translate the above considerations to
the disordered potential. In this case as well, we recognize that the
expansion of a BI in presence of a two-body interaction can be
described in terms of the dynamics of a two-fluid system, constituted
by local pairs (doublons) and unpaired particles. We are thus lead to
infer that the Anderson localization scenario must remain valid,
separately, for doublons and unpaired particles.  Furthermore, we can
determine how unpaired particles and doublons localize with different
localization lengths $l_{u}$ and $l_{d}$, the former being directly
given by \eqref{eq:1}, while  the latter corresponds the
the localization of a (composite) free particle whose dynamics is
regulated by the effective hopping $J^d=2J^2/\left|U\right|$, and by a
local potential $V^d_i=2 V_i$, namely
\begin{align}
 \label{eq:3a}
  l_u&=l_{loc}\left(\frac{V_0}{J}\right) \\
  \label{eq:3b}
  l_d&=l_{loc}\left(\frac{V^d_0}{J^d}\right)=l_{loc}\left(\frac{V_0 \left|U\right|}{J}\right).
\end{align}
The numerical simulations of the exact many-body dynamics show the
validity of the picture described above, according to which doublons
and unpaired particles localize independently following the AL
scenario.
In Fig. \ref{fig:3}a. we have plotted the localization properties for
three different setups ($U=0$, $V_0/J=5$, green curve; $U/J=-5$,
$V_0/J=1$, red curve; $U/J=-10$,$V_0/J=0.5$, blue curve). The
numerical results show that the three setups exhibit almost identical
localization lengths for unpaired paritcles ($U=0$, $V_0/J=5$) and
doublons ($U/J=-5$, $V_0/J=1$; $U/J=-10$,$V_0/J=0.5$). These numerical
results, in turn, are in agreement with the theoretical prediction
given by Eqs.~(\ref{eq:3a},\ref{eq:3b}), showing how the role of the
interaction profoundly modifies the localization scenario, and that
the correct identification of the elementary excitations in the system
represents a crucial step in the correct definition of the
localization properties of the system.  The deviation of the outer
tails from an exponential decay for $U/J=-10$, $V_0/J=0.5$ is related
to the slower propagation of the doublon wavefront, whose expansion
velocity is given by $\left|v_{d}\right|=4J^2/\left| U
\right|$. Therefore, at $t=60 J^{-1}$, the propagating doublon
wavefront is still visible in the plot (see Supplementary Information
- animation and \cite{Kajala:2011ho}). In Fig. \ref{fig:3}b. it is
possible to see the results of the exact numerical simulations
exhibiting two localization lengths for $U/J=-5$, $V_0/J=1$: the
central core is almost entirely constituted by doublons which
exponentially localize over a localization length $l_d=
l_{loc}\left(V_0 \left|U\right|/J\right)$, while the outer wings of
the expanding cloud are constituted by unpaired particles which
localize over a length $l_u=l_{loc}\left(\frac{V_0}{J}\right)$, in
agreement with the predictions given by the two-fluid model.

Another aspect of the localization properties of this system is given
by the independence of the doublon localization on the sign of the
interaction, which is connected to the unitarity of the evolution
considered here. The independence of $l_{loc}$ of the sign of the
interaction in Eqs.~(\ref{eq:3a},\ref{eq:3b} symmetry of the Hubbard
Hamiltonian under particle-hole transformation. The addition of a
(zero-average) disorder does not alter (in average) the dynamical
properties of the system, and thus allows to conclude that attractive
and repulsive interactions have the same effect in the localisation
properties of the system, as confirmed by the comparison of the
numerical results for $U/J=-5$,$V_0/J=1$ and $U/J=-5$,$V_0/J=1$ (see
Fig. \ref{fig:4}).
    
In conclusion, we have studied the expansion dynamics of an
interacting fermionic gase in a one-dimensional lattice, in presence of
on-site disorder. We have found that the two-body interaction affects
the localization properties of the system in a non-trivial way. We
have shown how the identification of the elementary excitations of the
system in terms of doublons and unpaired particles allows to recast
the rich localization scenarion observed in the numerical simulations
in terms of localization of these elementary excitations. Moreover, we
have shown that the localization length for this system does not
depend on the sign of the two-body interaction.

\textit{Acknowledgements.} The author would like to thank
A. Kupiainen, W. De Roek and F. Huveneers for useful discussions. 
This work was supported by ERC (ERC advanced grant MPOES). Computing
resources were provided by CSC - the Finnish IT Centre for Science.  

%\bibliographystyle{apsrev_abb}
%\bibliography{fh_al}

\end{document}